\documentclass{article}[11pt]
\usepackage{graphicx} 
\usepackage{amssymb,enumerate,graphicx,verbatim,natbib,amsthm,bbm}
\usepackage{todonotes}
\usepackage[inline, shortlabels]{enumitem}
\usepackage{thmtools}
\usepackage{thm-restate}
\usepackage[hidelinks]{hyperref}
\usepackage{subcaption}
\usepackage{mwe}
\usepackage[toc,page,header]{appendix}
\usepackage{minitoc}
\usepackage{algorithmic}
\usepackage[linesnumbered,ruled,vlined]{algorithm2e}
\usepackage[left=3.0cm, right=3.0cm, top=3cm]{geometry}
\usepackage{amsmath}
\usepackage[nameinlink,noabbrev,capitalise]{cleveref}
\Crefname{table}{Table}{Tables}
\Crefname{assumption}{Assumption}{Assumptions}
\crefname{equation}{}{}

\newtheorem{assumption}{Assumption}[section]

\usepackage{varwidth}
\usepackage{authblk}

\title{Treatment Effect Learning Under Sequential Randomization}

\author{Rina Friedberg}
\author{Richard Mudd}
\author{Patrick Johnstone}
\author{Melissa Pothen}
\author{Vishal Vaingankar}
\author{Vishwanath Sangale}
\author{Abbas Zaidi}

\affil{Meta Inc \authorcr
  \{\tt rinafriedberg, rmudd, austrartsua, melissapothen, vishalv, vishsangale, abbaszaidi\}@meta.com}
\date{}

\begin{document}

\maketitle

\section{Introduction}
Sequential treatment assignments in online experiments lead to complex dependency structures, often rendering identification, estimation and inference over treatments a challenge \citep{kohavi}. Treatments in one session (e.g., a user logging on) can have an effect that \textit{persists} into subsequent sessions leading to cumulative effects on outcomes (e.g., clicking on an ad) measured at a later stage.

\begin{figure}[h]
\centering
\includegraphics[scale=0.3]{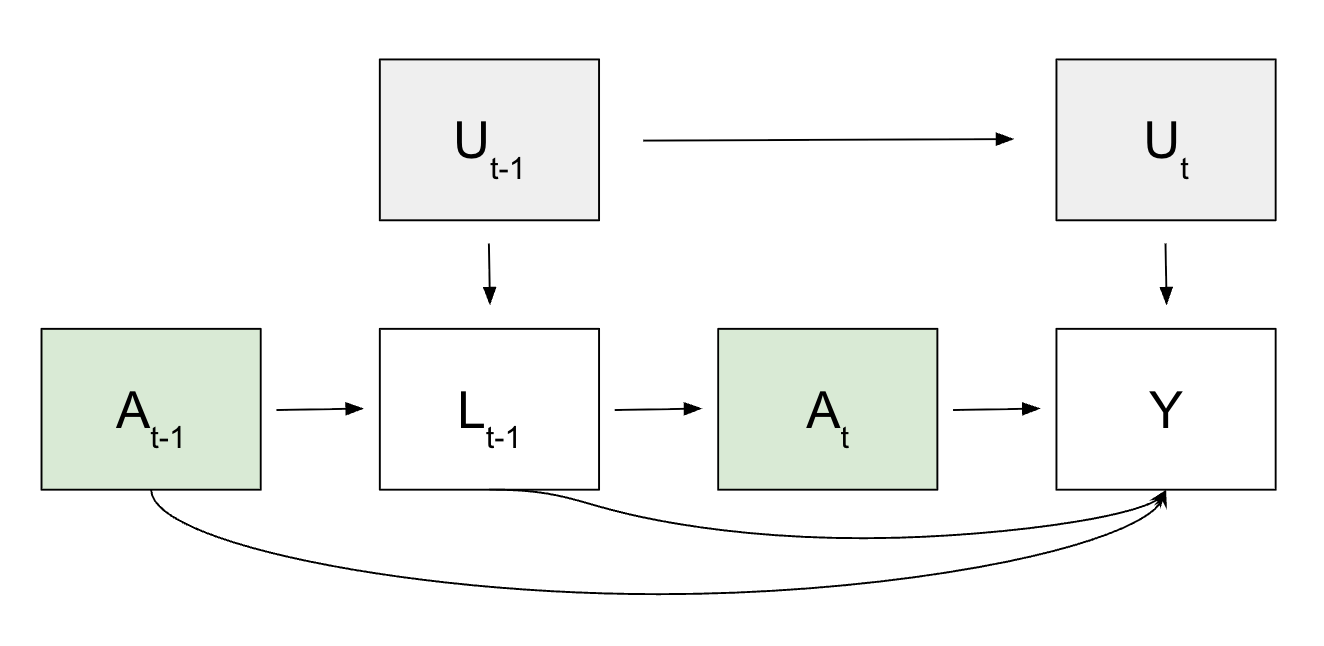}
\caption{DAG showing carry-over effects, confounding, and collider bias under sequential randomization with treatments $A_s$, lagged outcomes $L_s$, outcome $Y$, and unobserved confounding variables $U_s$.}
\label{fig:dag}
\end{figure}

Identification in sequential settings (Figure \ref{fig:dag}) entails thoughtful treatment of intermediate assignments and observations. Ignoring the effect of the initial treatment assignment $A_{t-1}$ via intermediate outcomes (i.e., just ignoring $L_{t-1}$ in estimation) leads to partial identification. In contrast, conditioning on $L_{t-1}$ can lead to two types of problems: if no unmeasured confounding is present, then any effect of the treatment through the intermediate outcome is partially blocked. In the presence of unmeasured confounding, \textit{collider bias} \citep{elwert2014} can occur: a well-known challenge in causal inference \citep{NguyenDafoeOgburn, hernan2017, holmberg2022}, wherein the treatment and the unmeasured confounder are common causes of the same third variable, here $L_t$ - a collider. Hence by conditioning on this intermediate outcome we can no longer identify the effect. 

To avoid said pitfalls we must first relax the conditional exchangeability assumption in favor of sequential conditional exchangeability \citep{hernan2017}. Under this assumption, the class of G-methods \citep{robins2003general} (which includes the G-formula, along with marginal structural models and structural nested models) can be used to identify the counterfactual and thereby enable causal conclusions.

There is a rich literature in machine learning for causal inference (among many; \citet{imai2013, athey2016recursive, chernozhukov2018double, hill2020}). Many such model classes are not designed to reflect the sequential nature of the assignment mechanism, leading to biased inferences as a consequence when not appropriately adjusted for.

We focus on the marriage of the G-formula which enables non-parametric identification of causal effects under sequential conditional exchangeability -- with machine learning approaches to enable causal inference at scale \citep{naimi2017intro}. In this work, we build on meta-learners (e.g., T-Learners, X-Learners) \citep{kunzel2019meta} within the G-formula as appropriate mechanism for causal identification under sequential conditional exchangeability. The remainder of this paper is organized as follows: section \ref{sec:methods} characterizes the methods, with results in section \ref{sec:results}, and discussion on areas of future investment in section \ref{sec:discussion}.

\section{Methods}
\label{sec:methods}
Suppose we observe a vector of treatments $\vec{a}$ taking values in $\{0,1\}$ and an outcome $Y = Y_T$. 
Suppose further that we have lagged outcomes or correlated measures $L_t$ measured at each time.
The target of inference is the treatment effect of $\vec{a}$ compared to a different vector of assignments $\vec{a}'$, that is, \begin{equation*} \tau_{\vec{a},\vec{a}'} = \mathbb{E}[Y(\vec{a})] - \mathbb{E}[Y(\vec{a}')]
\end{equation*}

We formalize the assumptions necessary for the identification of the above as mentioned in the previous section.

\subsection{Assumptions}

Identifying a treatment effect requires rests upon a series of assumptions -- The assumptions here are generalizations of standard causal inference assumptions to account for the complicated dependency structure at play \citep{naimi2017intro} in the sequential setting

\begin{assumption}[Sequential conditional exchangeability]\label{assn-seq-exch} At each time point $t$, and for any value of $L_{t-1}$, 
\begin{align}
    \mathbb{E}[Y(a_t, a_{t-1}) | A_t, A_{t-1}, L_{t-1}]  &=  \mathbb{E}[Y(a_t, a_{t-1}) | A_{t-1}, L_{t-1}], \text{~and} \\
\mathbb{E}[Y(a_t, a_{t-1}) | A_{t-1}]  &=  \mathbb{E}[Y(a_t, a_{t-1})]. \nonumber
\end{align}
\end{assumption}

\begin{assumption}[Positivity] For the first session, we must assume that $0 < P(A_0 =1) <1$. For every subsequent time point t and possible lagged outcome value $l_{t-1}$, we assume
\begin{equation}
    0 < P[A_t = 1 | L_{t-1} = l_{t-1}, A_{t-1} = a_{t-1}]  < 1
\end{equation}
\end{assumption}

\noindent While not required to identify a causal effect, the following simplifies the eventual estimation under the identification strategy characterized in this paper.

\begin{assumption}
The information provided from time T-1 encompasses the information provided from times 1 through T-1; that is, $\mathbb{E}[Y(\vec{a})] = \mathbb{E}[Y(\vec{a}) | A_T, A_{T-1}, L_{T-1}]$.
\end{assumption}

\subsection{Algorithm}
For each sequence $\vec{a}$ of interest, identification of the potential outcomes comes by way of leveraging various outcome modeling techniques within the G-formula in equation \ref{eq_xg},

\begin{equation}\label{eq_xg}
\mathbb{E}[Y(\vec{a})] = \sum_{\ell}\mathbb{E}[Y = y| L_1 = \ell, A_1 = a_1, A_0 = a_0] P [L_1 = \ell \mid A_0 = a_0]
\end{equation}

To estimate $\mathbb{E}[Y(\vec{a})]$, as a virtue of non-parametric identification under the G-formula, we specify models for each outcome on the left-hand-side of \ref{eq_xg}. Under the aforementioned outcome models, we will leverage Monte Carlo based estimation as in  algorithm \ref{alg_xg} for categorical (or reasonably bucketed) outcomes $L$ and $Y$; if these must be treated as continuous, we replace the sums with integrals.

 \begin{algorithm}
 Estimate an outcome model $g(Y\mid L_1,A_0,A_1) = \hat{\mathbb{E}}[Y \mid L_1, A_0, A_1]$ over $n$ units\\
 Estimate an outcome model $f(L_1\mid A_0) = \hat{P}[L_1 \mid A_0]$ over $n$ units\\
%
 \For{any sequence $\vec{a} = (a_0, a_1)$, over $n$}{
     \For{$k \in 1\dots, K$}{
        Simulate $\ell_1^{(k)}$ from $f(L_1\mid a_0)$. \\
        Estimate $\hat{P}^{(k)}[L_1 \mid A_0 = a_0]$ with $\sum_{n} \hat{P}[L_1 = \ell_1^{(k)} \mid A_0 = a_0] 1\{L_1 = \ell_1^{(k)}\}$. \\
        Simulate $y^{(k)}$ from $g(Y\mid\ell^{(k)},a_0,a_1)$.\\

    Estimate $\hat{\mathbb{E}}^{(k)}[Y\mid L_1 = \ell_1^{(k)}, A_1 = a_1, A_0 = a_0]$ with  $\sum_{n} y \hat{P}[Y = y \mid L_1 = \ell_1^{(k)}, A_1 = a_1, A_0 = a_0]$. \\
    
    Estimate  $\hat{\mathbb{E}}^{(k)}[Y(\vec{a})] = \sum_\ell \hat{\mathbb{E}}^{(k)}[Y\mid L_1 = \ell_1^{(k)}, A_1 = a_1, A_0 = a_0] \hat{P}[L_1 = \ell \mid A_0 = a_0]$.\\
    }
    
    Estimate $\hat{\tau}_{\vec{a},\vec{a}'} = \frac{1}{K}\hat{\mathbb{E}}^{(k)}[Y(\vec{a})] - \hat{\mathbb{E}}^{(k)}[Y(\vec{a}')]$.
 }
\caption{Sequential Estimation within the G-Formula.}
\label{alg_xg}
\end{algorithm}

Note that when all conditional distributions in g-computation are correctly specified, this approach leads to the most efficient estimates with the smallest large sample variances.

\subsection{Analysis Framework}
To operationalize algorithm \ref{eq_xg}, we specify T-learner base models for each of the outcome on the left hand side of equation \ref{eq_xg}. The choice of this type of flexible model is motivated by a key concern in the deployment of this strategy at scale: the \textit{g-null paradox} \citep{mcgrath2022revisiting}.

Under this setting, model misspecification leads to hypothesis tests that inevitably reject the null as sample size increases even when the causal null hypothesis is true. As this risk cannot be easily assessed by frequentist approaches, we attempt to guardrail against misspecification via expanding the model space with flexible approaches (e.g., the T-Learner) as overly parsimonious models exacerbate this risk.

With the base models specified, simulation for estimation assumes a Gaussian underlying distribution as in \citet{kunzel2019meta} for both the intermediate and final outcome stages. 

\subsection{Sensitivity Analysis}

While we use non-parametric models as part of our empirically calibrated simulation one could alternatively use more stringent parametric models where scale is a concern. However, as discussed in \citet{mcgrath2022revisiting}, because parametric  model misspecification may introduce bias in g-formula estimates, it is important to avoid overly parsimonious representations with appropriate sensitivity analysis mechanisms to guard against the g-null paradox. 

Such analyses are useful guardrails even with more flexible models such as those used in this work, where they function as additional protections against false rejections due to misspecification. In this example, appropriate sensitivity analyses such as those discussed in \citet{liu2013} would assess the bias of estimates derived from equation \ref{eq_xg} under misspecified models.

\section{Results}
\label{sec:results}
To demonstrate the effectiveness we offer a short simulation study calibrated against experiments that we have observed in practice over the the course of 2025. Suppose we have a sequential experiment set up with main treatment effects $\delta$ and carry-over effects $\eta$ across two sessions. We introduce no additional noise in order to highlight the identification strategy (as opposed to estimation or inference) that is, we simply model $Y = \delta a_1 + \eta a_0$.

We compare the mean squared error (`MSE') of $\tau_{(1,1), (0,0)}$ against the estimation target of $\delta + \eta$ between an T-Learner and T-Learner within the G-formula in Table \ref{tab_results} -- this choice is motivated by applications where T-Learners are the status quo. When carry-over effects exists, the MSE of the T-Learner degrades substantially – while T-learner used within the G-formulas, which are designed to accommodate this behavior, remain robust and correctly identify the treatment effect.

\begin{table}[h!]
\begin{subtable}{1\linewidth}
        \centering 
        \begin{tabular}{|c|c|c|c|}
            \hline
            T-Learner& $\delta$ & $\eta$ & MSE \\
            \hline
            &-0.217 & +0.055 & 0.050 \\
            &+0.118 & -0.015 & 0.050 \\
            &-0.402 & +0.023 & 0.030 \\
            \hline
        \end{tabular}
        \vspace{2mm}
        \centering
        
        \begin{tabular}{|c|c|c|c|}
            \hline
            T-Learner with G-Formula & $\delta$ & $\eta$ & 
            MSE \\
            \hline
            &-0.217 & +0.055 & 0.040 \\
            &+0.118 & -0.015 & 0.008 \\
            &-0.402 & +0.023 & 0.009 \\
            \hline
        \end{tabular}
\end{subtable}
\caption{Comparison of Mean Squared Error (/1000) between an T-Learner (top) and Using the T-Learner as a base-learner within the G-Formula (bottom)}
\label{tab_results}
\end{table}

Further, as the size of the carry-over effect increases in magnitude relative to the main treatment effect the performance of the model that fails to price them in degrades substantially. Finally, simulations that are not presented here for brevity  demonstrate that when no carry-over effects are present, the two approaches perform comparably as the identification assumptions underlying the T-learner hold.

\section{Discussion}
\label{sec:discussion}
Sequential assignments in experiments can render standard methods for identification trivially misspecified making estimation and inference moot. We propose layering meta-learners into the G-Formula for this setting, building on literature from causal machine learning and identification in sequential settings. 
Within a simple, empirically calibrated simulation, this approach prevents decaying accuracy gauged via the mean-square error in the presence of carry-over effects, highlighting the importance of identification strategies tailored to the nature of systems often seen in the tech domain.

Our proposed approach offers several areas for potential future research specifically for estimation and inference drawing from \citep{bang2005doubly}. Our choice of model specification for instance attempts to guard against the g-null paradox via model flexibility -- however characterizing this problem within the Bayesian paradigm means that the g-null paradox can be assessed by examining whether the prior predictive distribution of the
potential outcomes rules out the g-null hypothesis \citep{keil2018bayesian} offering a more direct approach to mitigating this risk. Further, introducing information via the prior offers a meaningful step in improving the accuracy of this system via tackling selection, particularly in under-powered settings \citep{kessler2024overcoming}. Finally, as inference remains a key component of business decisions, formally studying uncertainty quantification of these approaches is necessary. We aim to cover these methods in future research.

\bibliographystyle{abbrvnat}
\bibliography{references}

\end{document}